\documentstyle[prl,aps,epsfig]{revtex}

\begin{document}
%
%
\def\ov{\over}
\def\l{\left}
\def\r{\right}
\def\be{\begin{equation}}
\def\ee{\end{equation}}
\def \der#1#2{{\partial{#1}\over\partial{#2}}}
\def \dder#1#2{{\partial^2{#1}\over\partial{#2}^2}}
\def\R{{\rm I\! R}}
\def\spose#1{\hbox to 0pt{#1\hss}}
\def\lta{\mathrel{\spose{\lower 3pt\hbox{$\mathchar"218$}}
     \raise 2.0pt\hbox{$\mathchar"13C$}}}
\def\gta{\mathrel{\spose{\lower 3pt\hbox{$\mathchar"218$}}
     \raise 2.0pt\hbox{$\mathchar"13E$}}}
\def\ulap{\underline\Delta}
\def\unab{{\overline\nabla}}
\def\bisn{{\displaystyle B^i \over N}}
\draft
\title{Numerical models of irrotational binary neutron stars in general
relativity} 
\author{Silvano Bonazzola, Eric Gourgoulhon and Jean-Alain Marck}
\address{D\'epartement d'Astrophysique Relativiste et de Cosmologie \\
  UPR 176 du C.N.R.S., Observatoire de Paris, \\
  F-92195 Meudon Cedex, France \\
  {\sl e-mail: Silvano.Bonazzola, Eric.Gourgoulhon,
        Jean-Alain.Marck@obspm.fr}
}
\date{10 December 1998}
\maketitle

\begin{abstract} 
We report on general relativistic calculations of
quasiequilibrium configurations of binary neutron stars in circular
orbits with zero vorticity.  These configurations are expected to
represent realistic situations as opposed to corotating
configurations.  The Einstein equations are solved under the assumption
of a conformally flat spatial 3-metric (Wilson-Mathews approximation).
The velocity field inside the stars is computed by solving an
elliptical equation for the velocity scalar potential.  Results are
presented for sequences of constant baryon number (evolutionary
sequences). Although the central density decreases much less with the
binary separation than in the corotating case, it still decreases.
Thus, no tendency is found for the stars to individually collapse to
black hole prior to merger.  
\end{abstract}


Inspiraling neutron star binaries are expected to be among the
strongest sources of gravitational radiation that could be detected by
the interferometric detectors currently under construction (GEO600,
LIGO, TAMA and Virgo).  These binary systems are therefore subject to
numerous theoretical studies.  Among them are fully relativistic
hydrodynamical treatments, pioneered by the works of Oohara and
Nakamura (see e.g.  \cite{OoharN97}) and Wilson et al.
\cite{WilsoM95,WilsoMM96}. The most recent numerical calculations,
those of Baumgarte et al.  \cite{BaumgCSST97,BaumgCSST98} and
Marronetti et al. \cite{MarronMW98}, rely on the approximations of (i)
a quasiequilibrium state and (ii) of synchronized binaries.  Whereas
the first approximation is well justified before the innermost stable
orbit, the second one does not correspond to physical situations, since
it has been shown that the gravitational-radiation driven evolution is
too rapid for the viscous forces to synchronize the spin of each
neutron star with the orbit \cite{Kocha92,BildsC92} as they do for
ordinary stellar binaries.  Rather, the viscosity is negligible and 
the fluid velocity circulation (with respect to some inertial
frame) is conserved in these systems.  Provided that the initial spins
are not in the millisecond regime, this means that close binary
configurations are well approximated by zero vorticity (i.e. {\em
irrotational}) states.

Moreover, dynamical calculations by Wilson et al.
\cite{WilsoM95,WilsoMM96} indicate that the neutron stars may
individually collapse into a black hole prior to merger. This
unexpected result has been called into question by a number of authors
(see Ref.~\cite{MatheMW98} for a summary of all the criticisms and
their answers).  As argued by Mathews et al.~\cite{MatheMW98}, one way
to settle this crucial point is to perform computations of relativistic
irrotational configurations. We present here the first
quasiequilibrium irrotational relativistic binary neutron stars
models computations.

We have proposed a relativistic formulation for quasiequilibrium
irrotational binaries \cite{BonazGM97b} as a generalization of the
Newtonian formulation presented in Ref.~\cite{BonazGHM92}.  The method was
based on one aspect of irrotational motion, namely the {\em
counter-rotation} (as measured in the co-orbiting frame) of the fluid
with respect to the orbital motion (see also Ref.~\cite{Asada98}).
Since then Teukolsky~\cite{Teuko98} and Shibata~\cite{Shiba98} gave two
formulations based on the definition of irrotationality, which implies
that the specific enthalpy times the fluid 4-velocity is the gradient
of some scalar field \cite{LandaL89} ({\em potential flow}).  The three
formulations are equivalent; however the one given by Teukolsky and by
Shibata greatly simplifies the problem.

The hydrodynamic equations may be derived as follows. For a perfect
fluid at zero temperature, the momentum-energy
conservation equation $\nabla\cdot{\bf T}=0$ is equivalent to the 
{\sl uniformly canonical equation of motion}\cite{Lichn67,Carte79}, 
\be
\label{e:ucem}
{\bf u} \cdot (\nabla \wedge {\bf w}) = 0 
\ee
and
\be
\label{e:conserv}
\nabla \cdot (n {\bf u}) = 0 \ ,
\ee
where $\bf u$ is the fluid 4-velocity and $\bf w$ 
the momentum density 1-form ${\bf w} = h {\bf u}$, 
$h$ being the fluid specific enthalpy $h = (e+p) / (m_{\rm B} n)$ 
($\nabla \wedge {\bf w}$  denotes the exterior derivative of {\bf w}).
In the above equations, $n$, $e$ and $p$ denote respectively the 
fluid proper
baryon density, proper total energy density and pressure. 
It is clear that a potential flow 
\be \label{e:potf}
{\bf w} = \nabla \Psi
\ee
is a solution of the equation of motion (\ref{e:ucem}). Moreover, this 
particular solution is the relativistic generalization of the classical 
irrotational flow and, as stated above,
corresponds to the physical situation reached by a 
binary system of neutron stars.

As a first approximation of the relativistic treatment of the problem,
we will assume that there exists a helicoidal symmetry
\cite{BonazGM97b}. Let us denote by {\bf l} the associated Killing vector.
It is to be noticed that this symmetry is
exact at the Newtonian limit.  The helicoidal symmetry implies
$ {\cal L}_{\bf l} {\bf w} = 0$.
From Cartan's identity
${\cal L}_{\bf l} {\bf w} = {\bf l} \cdot \nabla \wedge {\bf w} 
        + \nabla ({\bf l} \cdot {\bf w}) $,
the potential form (\ref{e:potf}) leads immediately to the following 
first integral 
of motion
\be \label{e:int1}
{\bf l} \cdot {\bf w} = {\rm Const} \ .
\ee
This was first pointed out by Carter \cite{Carte79}. Note that this
result is not merely the relativistic generalization of the Bernoulli
theorem which states that ${\bf l} \cdot {\bf w}$ is constant along
each single streamline and which results directly from the existence of
a Killing vector without any hypothesis on the flow. In order for the
constant to be uniform over the streamlines (i.e. to be a constant over
spacetime), as in Eq.~(\ref{e:int1}), some additional property of the
flow must be required. One well known possibility is rigidity (i.e.
$\bf u$ colinear to $\bf l$) \cite{Boyer65}. The alternative property
with which we are concerned here is irrotationality.

The fluid motion is now completely determined by the scalar potential 
$\Psi$. The equation for $\Psi$ can be derived from 
the baryon number conservation 
(\ref{e:conserv}). One obtains:
\be
{n \over h} \nabla \cdot \nabla \Psi 
        + \nabla \Psi \cdot \nabla \left( {n \over h} \right) =0 \ .
\ee
Within the 3+1 formalism and taking into account the helicoidal symmetry, 
this last equation is nothing but a Poisson-like equation which 
reads\footnote{Latin indices run in the range 1,2,3 and geometrized units
($G=1$ and $c=1$) are used.}
\be \label{e:psicov}
n D_i D^i \Psi + D^i n D_i \Psi 
  = - {h \Gamma_{\rm n} \over N} B^i D_i n
  + n \left\{ \left( D^i \Psi + {h \Gamma_{\rm n} \over N} B^i \right) 
	D_i \ln h
     - D^i \Psi D_i \ln N - {B^i \over N} D_i (h \Gamma_{\rm n}) \right\}
  + K n h \Gamma_{\rm n}
\ ,
\ee
where we have introduced the covariant derivative $D$ with respect to
the spatial 3-metric, the trace $K$ of the extrinsic curvature tensor,
the shift vector $\bf B$ and the lapse $N$ defined by the 3+1
orthogonal decomposition of the helicoidal Killing vector ${\bf l} = N
{\bf n} - {\bf B}$ ($\bf n$ being the unit future directed normal
vector to the hypersurface $t = \rm Const$) and the Lorentz factor
$\Gamma_{\rm n} = - {\bf n} \cdot {\bf u}$ of the fluid with respect to the
Eulerian observer whose 4-velocity is $\bf n$. This equation has been
recently derived by Teukolsky \cite{Teuko98} and independently by
Shibata \cite{Shiba98}.  Note that Eq.~(\ref{e:psicov}) is independent
of the gravitational field equations.

As a first milestone in our project of studying coalescing binary 
systems, we will adopt the Wilson-Mathews 
approximation for the form of the metric \cite{WilsoM89,WilsoM95}.
This approximation consists in taking a conformally flat 3-metric, so that
the full spacetime metric reads
\be \label{e:metric}
ds^2 = -(N^2 - B_i B^i) dt^2 - 2 B_i dt\, dx^i 
        + A^2 \eta_{ij} dx^i\, dx^j \ ,
\ee
where {\boldmath $\eta$} is the flat space metric. The field equations
reduce now to the Wilson-Mathews equations \cite{WilsoM95,WilsoMM96} 
for $N$, $\bf B$ and $A$.
Right now, it is not obvious whether the Wilson-Mathews approximation
is valid in the case of coalescing compact binaries. We
presently use this particular form of the metric in order to simplify
the problem. However, it is to be noticed that (i) the 1-PN approximation
to Einstein equations fits this form, (ii) it is exact for arbitrary
relativistic spherical configurations and (iii) it is very accurate for
axisymmetric rotating neutron stars \cite{CookST96}.  An interesting discussion
about some justifications of the Wilson-Mathews approximation may be
found in \cite{MatheMW98}. Finally we chose maximal slicing: $K=0$. 

Following \cite{BonazGM97b}, we
introduce $\Omega$ such that the helicoidal 
Killing vector {\bf l} satisfies 
\be
{\bf l} = {\partial \over \partial \tau}
        + \Omega {\partial \over \partial \phi} \ , 
\ee
where $\tau$ and $\phi$ are respectively the time and azimuthal 
coordinate associated with the asymptotic inertial observer at rest
with respect to the binary system (i.e. such that the ADM 3-momentum
vanishes on the slices $\tau={\rm const}$). Besides, we introduce the {\sl 
non-rotating} shift vector $\bf N$ defined by
\be
{\bf B} = {\bf N} - \Omega {\partial \over \partial \phi} \ .
\ee

The gravitational field equations are derived within the 3+1 formalism 
from the Hamiltonian constraint, momentum constraint and trace of the spatial 
part of the Einstein equation \cite{WilsoMM96,BaumgCSST98}. 
Introducing $\nu = \ln N$ and $\beta = \ln (A N)$, they can be written as
\be \label{e:beta}
\ulap \beta = 4 \pi A^2 S + {3 \over 4} A^2 K_{ij} K^{ij} 
        - {1 \over 2} \left( \unab_i \nu \unab^i \nu  
        		+ \unab_i \beta \unab^i \beta \right) 
\ee
\be \label{e:nu}
\ulap \nu = 4 \pi A^2 (E + S) + A^2 K_{ij} K^{ij} - \unab_i \nu \unab^i \beta
\ee
\be \label{e:shift}
\ulap N^i + {1 \over 3} \unab^i \left( \unab_j N^j \right) =
        - 16 \pi N A^2 (E + p) U^i + 2 N A^2 K^{ij} \unab_j (3 \beta - 4 \nu)
	\ ,
\ee
where $\unab$ is the covariant derivative associated with the flat
3-metric {\boldmath $\eta$} and $\ulap = \unab^i \unab_i$ 
is the corresponding 
Laplacian\footnote{Throughout the article, we use the notation 
$\unab^i = \eta^{ij} \unab_j$.}.
$E = \Gamma_{\rm n}^2 (e+p) - p$, $S = 3 p + (E +p) U_i U^i$ and 
$U^i = D^i \Psi / (h \Gamma_{\rm n})$ 
are respectively the fluid energy density, trace of the stress tensor
and fluid 3-velocity, the three of them measured by the Eulerian observer.
$\Gamma_{\rm n}$ can be computed according to
\be
 \Gamma_{\rm n} = \l[ 1 + {1\ov A^2 h^2} 
                \unab_i \Psi \, \unab^i \Psi \r] ^{1/2}
        = {1 \over \sqrt{
                1 - U_i U^i
                }
        }
\ee
and the extrinsic curvature tensor is computed by means of the identity
\be \label{e:kij}
K^{ij} = - {1 \over 2 A^2 N}
          \left\{
                \unab^i N^j + \unab^j N^i
                  - {2 \over 3} \eta^{ij} \unab_k N^k 
          \right\} \ ,
\ee
which results from the Killing equation for $\bf l$.

The matter distribution is determined by the first integral of motion
(\ref{e:int1}). Taking its logarithm leads to
\be \label{e:int1_final}
  H + \nu +{1\ov 2}\ln\left( 1 - A^2 \eta_{ij} {B^i B^j\over N^2} \right)
        + \ln\Gamma = {\rm Const} \ ,
\ee
where $ H :=  \ln h$
and $\Gamma$ is the fluid Lorentz factor with respect to the co-orbiting
observer (i.e. the observer whose 4-velocity is collinear to $\bf l$):
\be
\Gamma = \Gamma_{\rm n} 
        { \displaystyle
        1 + A^2 \eta_{ij} {B^i \over N} U^j
        \over
        \sqrt{ \displaystyle
        1 - A^2 \eta_{ij} {B^i B^j \over N^2}
                }
        } \ . 
\ee
Note that Eq.~(\ref{e:int1_final}) corresponds to Eq.~(66) 
in Ref.~\cite{BonazGM97b} and that the constant which appears at its
r.h.s. is nothing but the logarithm of the constant $C$
introduced by Teukolsky \cite{Teuko98}.

\bigskip
Now, introducing $\zeta = d \ln H / d \ln n$, the equation for the fluid 
velocity potential (\ref{e:psicov}) becomes
\be \label{e:psinum}
\zeta H \ulap \Psi + \unab^i H \unab_i \Psi =
  - A^2 h \Gamma_{\rm n} \bisn \unab_i H
 + \zeta H \left\{
 \left(
        \unab^i \Psi + A^2 h \Gamma_{\rm n} \bisn \right) \unab_i H
        - \unab^i \Psi \unab_i \beta
        - A^2 \bisn \unab_i (h \Gamma_{\rm n})
                \right\} \ .
\ee

The equations to be solved (\ref{e:beta}) (\ref{e:nu}) (\ref{e:shift})
(\ref{e:psinum}) constitute a system of non-linear Poisson-like
equations. Because of the elliptical nature of these equations, we have
exhibited the common flat Laplacian operator $\ulap$ which can be solved by
means of usual spectral methods (cf. e.g. \cite{BonazGM97a} or
\cite{BonazGM98b}). Because of the non-linearities, we use an iterative
procedure based on a multi-domain spectral method \cite{BonazGM98a} to
get the solution. The numerical code will be described in
details in a forthcoming paper \cite{BonazGM99}. Let us simply mention here
some tests passed by the code. In the Newtonian and incompressible
limit, the analytical solution constituted by a Roche ellipsoid is
recovered with a relative accuracy of $\sim 10^{-9}$ (cf.  Fig.~6 of
Ref.~\cite{BonazGM98a}).  For compressible and irrotational Newtonian
binaries, no analytical solution is available, but the virial theorem
can be used to get an estimation of the numerical error: we found that the
virial theorem is satisfied with a relative accuracy of $10^{-7}$. A
detailed comparison with the irrotational Newtonian configurations
recently computed by Uryu \& Eriguchi \cite{UryuE98a,UryuE98b} will be
presented in \cite{BonazGM99}.  Regarding the relativistic case, we
have checked our procedure of resolution of the gravitational field
equations by comparison with the results of Baumgarte et
al.~\cite{BaumgCSST98} which deal with corotating binaries [our code
can compute corotating configurations by simply setting $\ln\Gamma =
0$ in Eq.~(\ref{e:int1_final}) and using $U^i= -B^i/N$ for the fluid
3-velocity]. We have performed the comparison with
the configuration $z_A=0.20$ in Table~V of Ref.~\cite{BaumgCSST98}. We
used the same equation of state (EOS) (polytrope with $\gamma=2$), same
separation $r_C$ and same value of the maximum density parameter
$q^{\rm max}$. We found a relative discrepancy of $1.1\%$ on $\Omega$,
$1.4\%$ on $M_0$, $1.1\%$ on $M$, $2.3\%$ on $J$, $0.8\%$ on $z_A$,
$0.4\%$ on $r_A$ and $0.07\%$ on $r_B$ (using the notations of
Ref.~\cite{BaumgCSST98}).

\begin{figure}
\centerline{\epsfig{figure=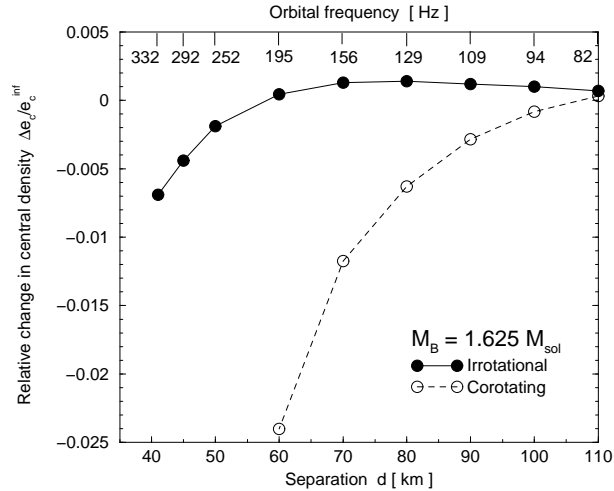,width=8cm}}
\caption[]{\label{f:dens} 
Relative variation of the central energy density $e_{\rm c}$ with
respect to its value at infinite separation $e_{\rm c}^{\rm inf}$ as a
function of the coordinate separation $d$ (or of the orbital frequency
$\Omega/(2\pi)$) for constant baryon mass $M_{\rm B} = 1.625 \,
M_\odot$ sequences.  The solid (resp. dashed) line corresponds to a
irrotational (resp. corotating) sequence of coalescing neutron star
binaries. Note that there is no substantial increase of the central
density as the separation decreases.}
\end{figure}

These tests being passed, we turned towards calculations of
irrotational relativistic binaries. We chose to investigate the
instability issue raised by Wilson and Mathews \cite{WilsoM95} by
computing an evolutionary sequence (i.e. a sequence at fixed baryon
number) made of irrotational quasiequilibrium models. We took the same
configuration than that presented by Mathews, Marronetti and Wilson
(Sect.~IV~A of Ref~\cite{MatheMW98}), namely two identical stars
obeying a $\gamma = 2$ polytropic EOS [$p=\kappa (m_{\rm B} n)^\gamma$,
$e=m_{\rm B} n + p/(\gamma-1)$] with $\kappa = 1.8 \times 10^{-2}
{\ \rm J\, m}^3{\rm kg}^{-2}$ and each having a baryon mass $M_{\rm B} =
1.625\, M_\odot$. For such parameters, we found that the gravitational
mass of a single star in isolation is $M=1.515\, M_\odot$ (in agreement
with Ref.~\cite{MatheMW98}), with a central energy density 
$e_{\rm c}^{\rm inf}=4.005\,\rho_{\rm nuc}\,c^2$ 
($\rho_{\rm nuc}:=1.66\times 10^{17}{\ \rm kg\,
m}^{-3}$) and a compactification ratio $M/R=0.140$ 
($e_{\rm c}$ and $M/R$ are
slightly different from that quoted in Ref.~\cite{MatheMW98} probably
due to different values for the constants $G$, $c$, $M_\odot$ and
$m_{\rm B}$; we use $G=6.6726\times 10^{-11} {\ \rm m}^3{\rm kg}^{-1}
{\rm s}^{-2}$, $c=2.99792458\times 10^8 {\ \rm m\, s}^{-1}$, $M_\odot =
1.989 \times 10^{30} {\rm \ kg}$ and $m_{\rm B}=1.66\times 10^{-27}{\rm
\ kg}$).  

We define the coordinate separation $d$ as the coordinate distance
between the two density maxima.  Using the same value as the one
considered by Mathews et al. \cite{MatheMW98}, namely $d=100
{\rm\ km}$, we found that a $M_{\rm B}=1.625\, M_\odot$ irrotational
configuration in quasiequilibrium at this separation has a total
angular momentum of $J/(2M_{\rm B})^2 = 1.13$ which is quite similar to
the value $1.09$ found by Mathews et al.  \cite{MatheMW98}. However,
we did not observed any substantial increase of the central (i.e. maximum)
density with respect to static stars in isolation, as they did 
(they report a central density increase of $14\%$).

In order to investigate the evolution of a coalescing binary system, we
have computed a full sequence at fixed $M_{\rm B}=1.625\, M_\odot$,
starting at the separation $d=110 {\rm \, km}$ 
($\Omega/(2\pi) = 82 {\rm\ Hz}$) and ending at $d=41{\rm
\, km}$ ($\Omega/(2\pi) = 332 {\rm\ Hz}$). 
We have considered both corotating and irrotational cases.
The evolution of the central density
along these sequences is shown in Fig.~\ref{f:dens}. 
In the corotating case, we found that
the central density decreases substantially when the stars approach
each other, as expected from previous independent calculations
\cite{BaumgCSST97,BaumgCSST98}. In the irrotational case, we found that
the central density still decreases with the separation but much less
than in the corotating case. 

\bigskip
All the calculations have been performed with spectral methods using the
library LORENE ({\sc Langage Objet pour la RElativit\'e Num\'eriquE} 
\cite{LORENE}). We warmly thank Jean-Pierre Lasota for his support.
The calculations presented in this paper have been made possible thanks
to a special grant from the SPM and SDU departments of CNRS.

\end{document}